\documentclass[aps,prl,twocolumn,showpacs,superscriptaddress]{revtex4}

\usepackage{graphicx}
\usepackage{amsmath}
\usepackage{amssymb}
\usepackage{epsfig}

\usepackage{ctable}
\usepackage{slashbox}

\renewcommand{\v}[1]{{\bf #1}}

\def\eqa{\begin{eqnarray}}
\def\eea{\end{eqnarray}}
\newcommand{\eq}{\begin{equation}}
\newcommand{\ee}{\end{equation}}
\newcommand{\nn}{\nonumber\\}

\newcommand{\p}{\partial}

\newcommand{\ua}{\uparrow}
\newcommand{\da}{\downarrow}
\newcommand{\ra}{\rightarrow}

\newcommand{\al}{\alpha}
\newcommand{\bt}{\beta}

\newcommand{\Del}{\Delta}
\newcommand{\eps}{\epsilon}

\newcommand{\ga}{\gamma}
\newcommand{\Ga}{\Gamma}
\newcommand{\ka}{\kappa}
\newcommand{\la}{\lambda}
\newcommand{\La}{\Lambda}

\renewcommand{\th}{\theta}

\newcommand{\si}{\sigma}

\newcommand{\sro}{ Sr$_2$RuO$_4$ }

\usepackage{color}
\begin{document}

\title{Theory of chiral $p$-wave superconductivity with near-nodes for Sr$_2$RuO$_4$}

\author{Wan-Sheng Wang}
\email{wangwansheng@nbu.edu.cn}
\affiliation{Department of Physics, Ningbo University, Ningbo 315211, China}
\affiliation{National Laboratory of Solid State Microstructures \& School of Physics,
	Nanjing University, Nanjing, 210093, China}

\author{Cong-Cong Zhang}
\affiliation{Department of Physics, Ningbo University, Ningbo 315211, China}

\author{Fu-Chun Zhang}
\affiliation{Kavli Institute for Theoretical Sciences \& CAS Center for Excellence in Topological Quantum Computation, University of Chinese Academy of Sciences, Beijing 100190, China}
\affiliation{Collaborative Innovation Center of Advanced Microstructures, Nanjing University, Nanjing 210093, China}

\author{Qiang-Hua Wang}
\email{qhwang@nju.edu.cn}
\affiliation{National Laboratory of Solid State Microstructures \& School of Physics,
	Nanjing University, Nanjing, 210093, China}
\affiliation{Collaborative Innovation Center of Advanced Microstructures, Nanjing University, Nanjing 210093, China}

\begin{abstract}

  We use functional renormalization group method to study a three-orbital model for superconducting Sr$_2$RuO$_4$. Although the pairing symmetry is found to be chiral $p$-wave, the atomic spin-orbit coupling induces near-nodes for quasiparticle excitations. Our theory explains a major experimental puzzle between $d$-wave-like feature observed in thermal experiments and the chiral $p$-wave triplet pairing revealed in nuclear-magnetic-resonance and Kerr effect.

\end{abstract}


%\pacs{74.20.-z}{Theories and models of superconducting state}
%\pacs{74.20.Pq}{Electronic structure calculations}
%\pacs{74.20.Rp}{Pairing symmetries (other than s-wave)}
%\pacs{: 74.20.-z, 74.25.Jb, 74.70.Dd}
%74.70.Xa Pnictides and chalcogenides
%75.30.Fv  Spin-density waves
%74.70.Wz  Carbon-based superconductors
%81.05.ue  Graphene
%73.22.Pr  Electronic structure of graphene
%74.20.Rp  Pairing symmetries (other than s-wave)
%74.20.-z  Theories and models of superconducting state
%71.27.+a  Strongly correlated electron systems; heavy Fermions
%64.60.ae  Renormalization-group theory
%71.10.-w Theories and models of many-electron systems
%74.62.Dh Effects of cystal defects, doping and substitution (Superconductivity)
%74.20.Mn : Nonconventional mechanisms
%74.25.Dw : Superconductivity phase diagrams
%74.70.-b :  Superconducting materials other than cuprates
%(for cuprates, see 74.72.-h; for superconducting films, see 74.78.-w)

\maketitle

{\bf Introduction}: The layered perovskite ruthenate \sro is one of the rare candidate materials that is expected to carry chiral $p$-wave pairing in the superconducting (SC) state. Nuclear magnetic resonance (NMR) \cite{ksO1, ksru1, ksru2, kso2} and spin polarized neutron scattering \cite{PNeutron} measurements show absence of drop in the spin susceptibility below the SC transition temperature $T_c$, providing identification of spin-triplet pairing in \sro. Muon spin relaxation \cite{uSR} and polar Kerr effect \cite{kerr} experiments reveal that time reversal symmetry in \sro is spontaneously broken below $T_c$, suggesting chiral $p$-wave triplet pairing. The $d$-vector of the triplet is proposed as $\hat{z} ( k_x \pm i k_y )$,\cite{rmp, Kallin} which is analogous to that in the superfluid $^3$He-A phase. \cite{rice95} In this case, the SC state is likely fully gapped, since no symmetry forces the chiral $p$-wave gap function to vanish on the quasi two-dimensional Fermi surface (FS) of the layered Sr$_2$RuO$_4$.

In experiments, however, low-energy quasi-particle excitations deep in the SC state, characteristic of gap nodes on the FS (forming nodal lines along the direction perpendicular to the RuO$_2$ plane), are observed in specific heat, \cite{cve1, cve2, cve3} superfluid density, \cite{sfde} spin-lattice relaxation rate, \cite{t1te} thermal conductivity \cite{kappa1,kappa2,kappa3} and ultrasound attenuation \cite{sound} at low temperatures. To explain the nodal-like behavior, a simple scenario is to assume $d$-wave pairing symmetry, so that the gap nodes are symmetry protected. This scenario is, however, inconsistent with the compelling signatures of chiral $p$-wave triplet mentioned above. An alternative scenario is the chiral $p$-wave gap function may have deep minima or accidental nodes.\cite{rice-hl,Nomura,Kivelson,Miyake,QHWepl} The linear specific heat and thermal conductivity below $T_c / 2$ suggest that the gap minimum $\Del_{\rm min}$ should be much smaller than the gap maximum $\Del_{\rm max}$.\cite{Nomura,Miyake} The recent thermal conductivity measurement \cite{kappa3} sets an upper bound $\Del_{\rm min}/\Del_{\rm max}\leq 1/100$, and the field dependence suggests $d$-wave pairing, or $d$-wave-like $f$-wave pairing in the form of $(k_x+ik_y)g(\v k)$, where $g(\v k)\sim k_x k_y$ or $k_x^2-k_y^2$.\cite{fwave1,fwave2,fwave3}

Sr$_2$RuO$_4$ has three energy bands ($\al$, $\bt$ and $\ga$, derived from the $d_{xz,yz,xy}$ orbitals) crossed by the Fermi level, with the $\ga$ Fermi pocket closer to the van Hove singularity (vHS) on the zone boundary. The singular-mode functional renormalization group (SMFRG) study of the three-orbital model without spin-orbit coupling (SOC) \cite{QHWepl} showed that the gap function on the $\ga$ pocket is largest and strongly anisotropic, with $\Del_{\rm min}/\Del_{\rm max}\sim 1/10$. However, such a gap structure is not yet enough to explain the linear specific heat and thermal conductivity at the measured low temperatures. Models with SOC were previously studied by using weak coupling RG and random phase approximation.\cite{Scaffidi, YaoHong} But to our knowledge, close and systematic comparisons to experiments have not been reported.

The outstanding puzzle of the chiral $p$-wave pairing revealed in NMR and Kerr effect and the $d$-wave-like behavior indicated in thermal experiments motivates us to perform more careful microscopic investigations.
We consider a comprehensive model including all of the three orbitals and the atomic SOC.\cite{Damascelli1, Damascelli2, Damascelli3, Damascelli4} We adopt the band structure (with the effect of SOC) that best fits the angular-resolved photo-emission spectroscopy measurement.\cite{tbm0} We apply the spin-resolved version of SMFRG \cite{xyy1, yy, yy2, wws3} and treat all possible ordering tendencies on equal footing.\\

\begin{figure}
	\includegraphics[width=\columnwidth]{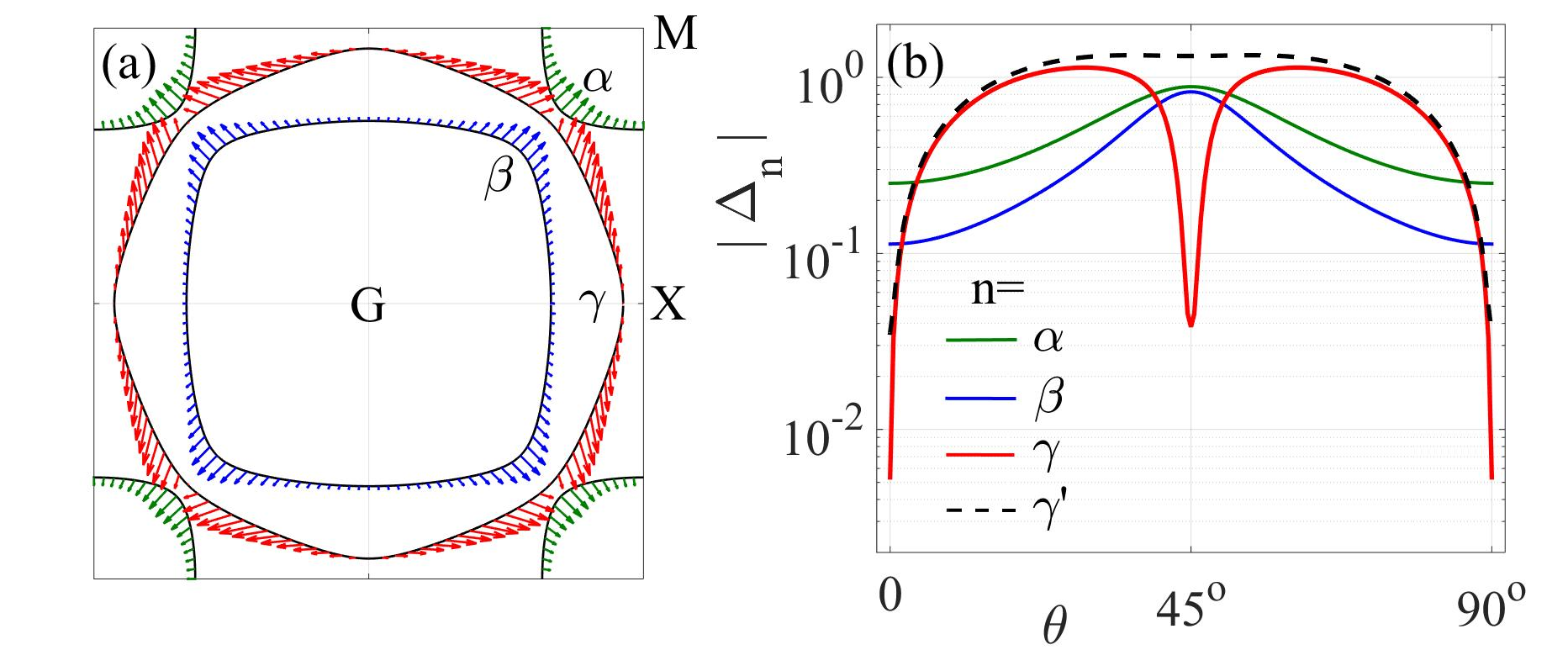}
	\caption{(a) FRG-derived $p_x + i p_y$-wave gap function on the FS (thin black lines). The arrow represents (Re $\Del_{\v kn}$, Im $\Del_{\v kn}$) for $n\in (\al,\bt,\ga)$. (b) The solid lines show the gap amplitude (up to a global scale) on the FS versus the Fermi angle $\th$ in a quadrant of the respective pocket. The dashed line shows the gap on the $\ga$ pocket if SOC is switched off artificially, showing the effect of SOC in generating deeper near-node along G-X ($\th=0$) and local minimum along G-M ($\th=45^{\rm o}$).}\label{fig:pip}
\end{figure}

\begin{figure}
	\includegraphics[width=\columnwidth]{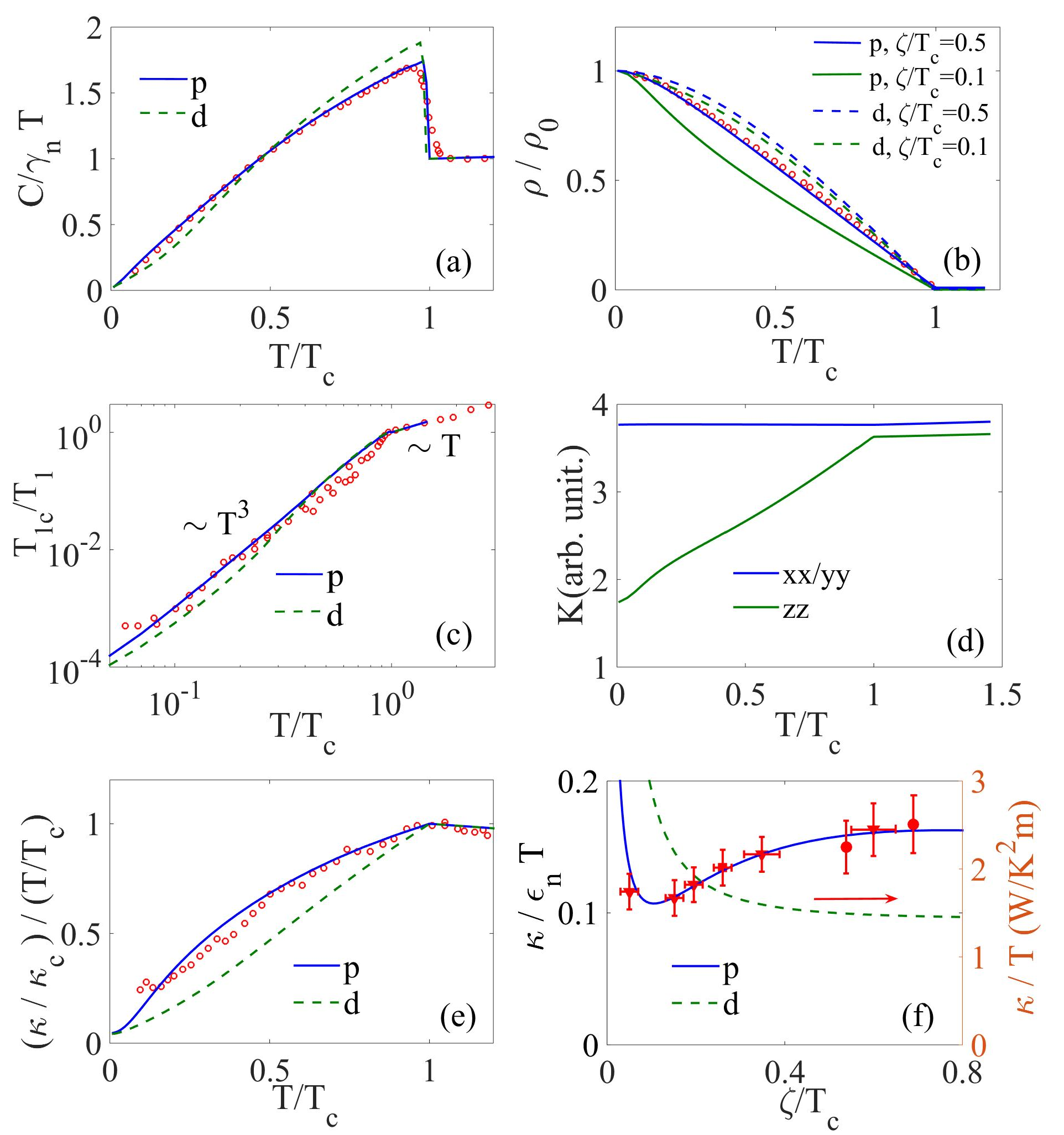}
	\caption{The calculated physical properties in the SC state (lines), in comparison with experimental data (symbols). The solid lines are for our chiral $p$-wave, while the dashed lines are $d$-wave fits.  (a) The electronic specific heat $C$ versus the temperature $T$. Here $\ga_n$ is the (constant) value of $C/T$ in the normal state. The symbols are extracted from Ref.\cite{cve1}, where $T_c=1.48K$. (b) Superfluid density $\rho$ versus $T$, with symbols from Ref.\cite{sfde}. (c) Spin-lattice relaxation rate $1/T_1$ versus $T$, normalized with respect to the value at $T_c$. The symbols are from Ref.\cite{t1te}, where $T_c = 1.48K$. (d) The direction-resolved Knight-shift $K$ versus $T$. (e) Thermal conductivity $\ka$ versus $T$. The symbols are from Ref.\cite{kappa3}. Here we use $\zeta/ T_c =0.26$ according to the experimental $T_c=1.2$ K. (f) Low temperature limit of $\ka/T$ as a function of impurity scattering rate $\zeta$. The temperature is fixed at $T_0=T_c/30$. The symbols are from Ref.\cite{kappa1} (circles), Ref.\cite{kappa2} (triangles) and Ref.\cite{kappa3} (squares). The numerical results are normalized by an assumed non-SC scale $\eps_n$, the value of $\ka/T$ at $T=T_0$, $\zeta=0.6T_c$, and zero gap.}\label{fig:prop}
\end{figure}

Our main results are summarized in Figs.\ref{fig:pip} and \ref{fig:prop}. We find that chiral $p$-wave pairing is dominant and can be related to the small-momentum spin fluctuations derived from the $d_{xy}$ orbital, similarly to the case in Ref.\cite{QHWepl}. However, SOC induces near-nodes on the $\ga$ pocket, with $\Del_{\rm min}/\Del_{\rm max}<1/100$. SOC also induces sizable and anisotropic gaps on the $\al$ and $\bt$ pockets. The calculated specific heat, superfluid density, Knight shift, spin-lattice relaxation rate and thermal conductivity are in excellent agreement with experimental data, superior to the $d$-wave fits. Our theory reconciles the $d$-wave-like feature in thermal measurement and chiral $p$-wave spin triplet pairing in NMR and Kerr effect in Sr$_2$RuO$_4$.\\

\begin{figure}
	\includegraphics[width=\columnwidth,clip]{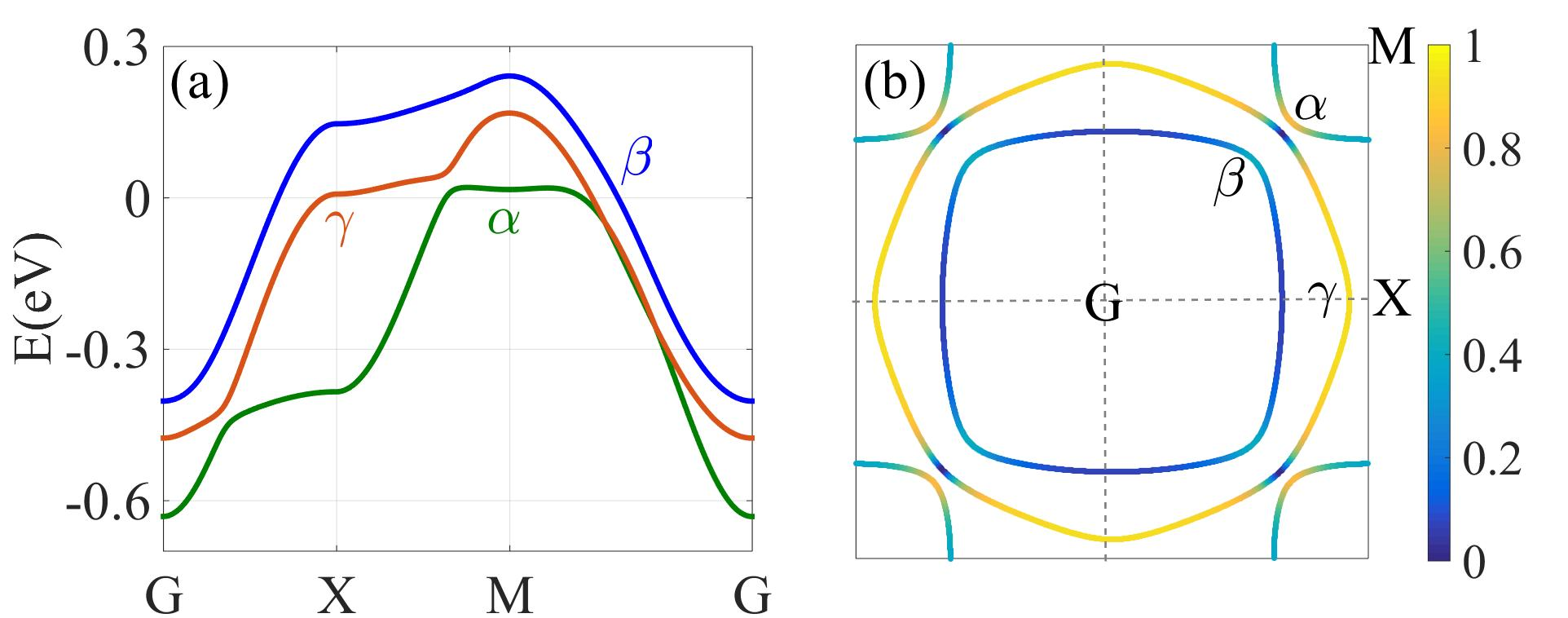}
	\caption{ (a) Band dispersion along high-symmetry cuts. (b) Fermi surface (lines) and the spectral weight of the $d_{xy}$-orbital (color scale) thereon.}\label{fig:band}
\end{figure}

{\bf Model and method}: We now specify the model Hamiltonian $H=H_0+H_I$ for Sr$_2$RuO$_4$. The free part can be written as
\eqa H_0 = \sum_\v k \psi_\v k^\dag h_\v k\psi_\v k,\ \ h_\v k=\eps_\v k\si_0-\la\v L\cdot\vec\si/2.\eea
Here $\psi_\v k =(c_{\v k 1\ua},c_{\v k 2\ua},c_{\v k 3\ua},c_{\v k 1\da},c_{\v k 2\da},c_{\v k 3\da})^T$ is the fermion spinor, with $c_{\v k a s}$ annihilating an electron of momentum $\v k$ and spin $s\in (\ua,\da)$ on orbital $a\in (1,2,3)\leftrightarrow (d_{xz},d_{yz}, d_{xy})$. In the single-particle Hamiltonian $h_\v k$, $\eps_\v k$ is a matrix in the orbital basis, $\v L$ is the orbital angular momentum, and $\vec\si/2$ is the spin angular momentum. The SOC parameter is $\la=0.032$ eV,\cite{tbm0} and the other details for $h_\v k$ can be found in Refs.\cite{tbm0,SM}
Fig.\ref{fig:band}(a) shows the band dispersion calculated with $H_0$ along high symmetry cuts. By inversion and time-reversal symmetries, each band is doubly degenerate in pseudo-spin.\cite{SM} Fig.\ref{fig:band}(b) shows the Fermi surface (FS). Note the $d_{xy}$-content of the Bloch state is dominant on the $\ga$ pocket, but vanishes identically along G-M.

\begin{figure}
	\includegraphics[width=0.75\columnwidth]{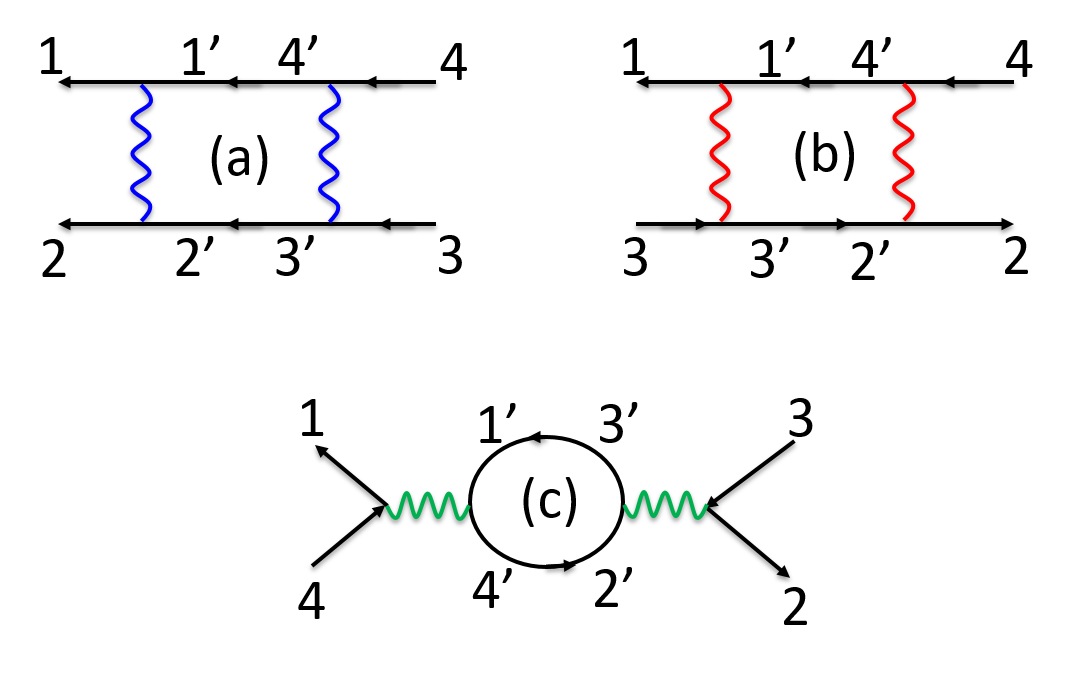}
	\caption{One-loop diagrams contributing to $\p\Ga_{1234}/\p\La$, quadratic in $\Ga$ itself (wavy lines, fully antisymmetrized with respect to incoming or outgoing fermions, labelled by the numerical indices). The color of the wavy line signifies the scattering of fermion bilinears in the pairing (blue), crossing (red) and direct (green) channels. }\label{fig:feynman}
\end{figure}

The interacting part of the Hamiltonian $H$ is given by, in real space,
%\begin{widetext}
\eqa H_{I}  &= &U \sum_{ia} n_{ia\ua}n_{ia\da}+J \sum_{i,a>b,ss'}c^{\dag}_{ia s}c_{ib s} c^{\dag}_{ibs'}c_{ias'} \nn
&+&U' \sum_{i,a>b} n_{ia}n_{ib}+J' \sum_{i,a\neq b}c^{\dag}_{ia\ua}c^{\dag}_{ia\da} c_{ib\da}c_{ib\ua}, \label{H} \eea
%\end{widetext}
where $i$ denotes the lattice site, $n_{ia}=\sum_{s} c_{ias}^\dag c_{ias}$, $U$ is the intra-orbital repulsion, $U'$ is the inter-orbital repulsion, $J$ is Hund's rule coupling, and $J'$ is the pair hopping term.
The interactions can lead to competing collective fluctuations in particle-hole (PH) and particle-particle (PP) channels, which we handle by SMFRG. Following the general idea of FRG, \cite{Wetterich} we obtain the one-particle-irreducible 4-point interaction vertices $\Ga_{1234}$ (where numerical index labels single-particle state) for quasi-particles above a running infrared energy cut off $\La$ (which we take as the lower limit of the continuous Matsubara frequency). Starting from $\La=\infty$ where $\Ga$ is specified by the bare parameters in $H_I$, the contribution to the flow (toward decreasing $\La$) of the vertex, $\p\Ga_{1234}/\p\La$, is illustrated in Fig.\ref{fig:feynman}. At each stage of the flow, we decompose $\Ga$ in terms of eigen scattering modes (separately) in the PP and PH channels to find the negative leading eigenvalue (NLE), the divergence of which signals an emerging order at the associated scattering momentum, with the internal microscopic structure described by the eigenfunction. The technical details can be found elsewhere,\cite{wws1, xyy1, xyy2,wws2, xyy3, yy, QHWepl, yy2, wws3, lyc} and also in Ref.\cite{SM}. \\

\begin{figure}
	\includegraphics[width=0.8\columnwidth]{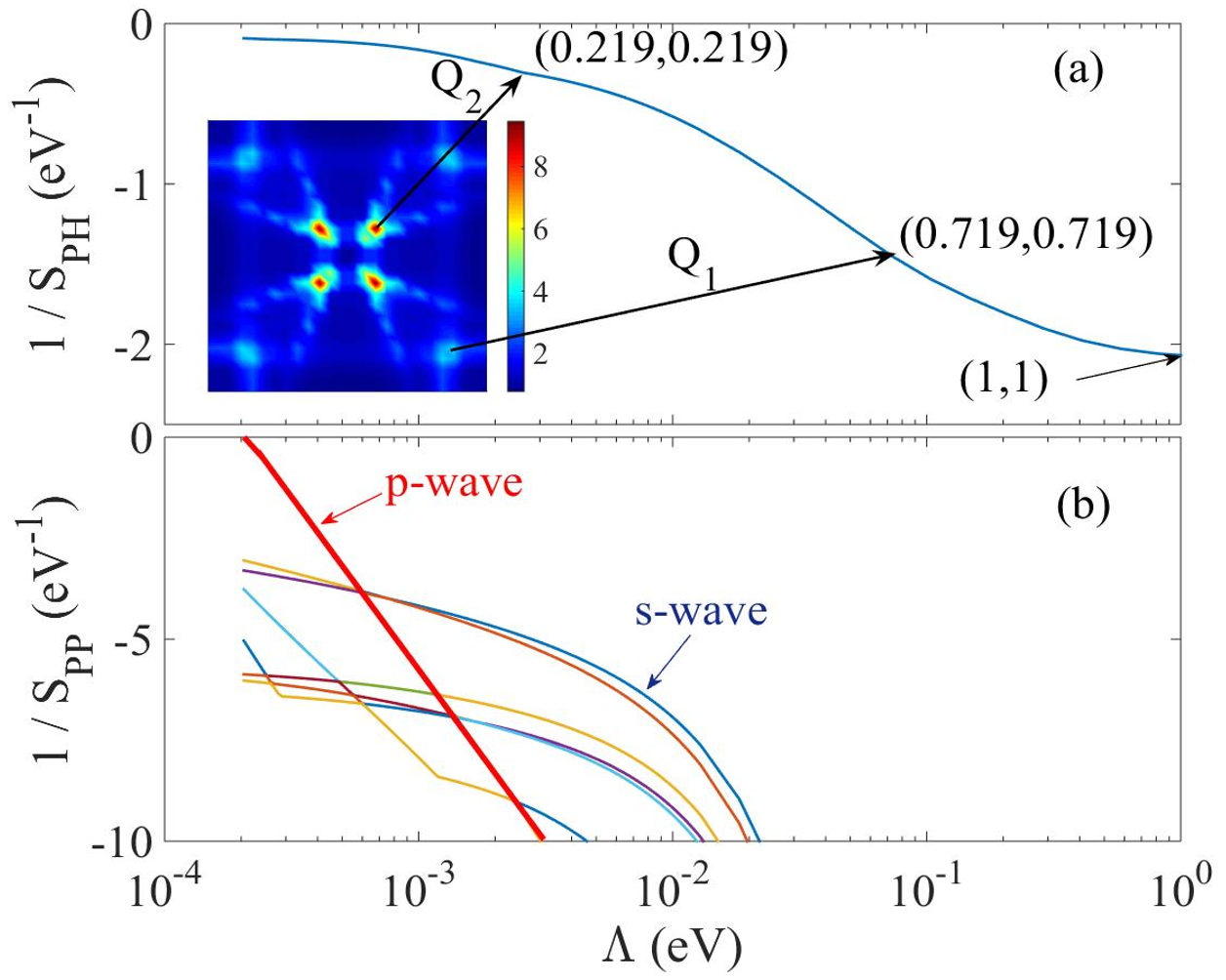}	
	\caption{(a) Flow of negative leading eigenvalue (among all momenta), $S_{\rm PH}$ in the PH channel, shown as $1/S_{\rm PH}$ for clarity. The inset shows $-S_{\rm PH}(\v q)$ in the momentum space at the divergence scale $\La=\La_c$. (b) Flow of NLE's $S_{\rm PP}(\v q=0)$. The thick line denotes the two eventually diverging $p-$wave pairing modes. Arrows indicate level crossing for $\v Q / \pi$ in the PH channel (a) and the pairing symmetries (b). }\label{frgf}
\end{figure}

{\bf Discussions}: We consider the bare interaction parameters $(U, U', J, J') = (0.4, 0.16, 0.04, 0.04)$ eV. The results are qualitatively robust against fine tuning of interactions and SOC around the present setting.\cite{SM} Fig.\ref{frgf}(a) shows the flow of NLE  $S_{\rm PH}$ (among all momenta) in the PH channel. The corresponding momentum changes from $\v Q_1\sim (0.719, 0.719)\pi$ at high energy scale to $\v Q_2 \sim (0.219,0.219)\pi$ in the intermediate stage. We checked that the eigenfunction  describes site-local spin. The $d_{xz,yz}$ ($d_{xy}$) orbitals dominate before (after) the level crossing. The inset shows the NLE $S_{\rm PH}(\v q)$ as a function of momentum $\v q$ at the final stage of the FRG flow. We see a strong peak at $\v Q_2$ and also a secondary peak at $\v Q_1$. These peaks are consistent with the spin-fluctuations observed in neutron scattering experiments.\cite{ins} Our results provide clear origins of such peaks: spin fluctuations at $\v Q_1$ ($\v Q_2$) arise mainly from the $d_{xz,yz}$ ($d_{xy}$) orbital, similarly to the case without SOC.\cite{QHWepl} At low energy scales, the PH channel saturates due to the decreasing phase space for low-energy PH excitations.

Fig.\ref{frgf}(b) shows ten NLE's in the PP channel (at zero momentum). They are induced at intermediate scales, where the PH channel is enhanced, a manifestation of the Khon-Luttinger mechanism,\cite{Luttinger} namely, the interaction in the PH channel has an overlap in the PP channel. Eventually, a particular mode (red thick line) diverges. We find it describes $p_{x,y}$-wave pairing (to be detailed below), and is twofold degenerate by $C_{4v}$ symmetry.
The details of the pairing function (the eigenfunction of the NLE scattering mode in the PP channel) are presented in Ref.\cite{SM} Here we show the projection of the $p_x+ip_y$ pairing function (favored in the SC state) in the band basis in Fig.\ref{fig:pip}. There are several remarkable features: (i) In Fig.\ref{fig:pip}(a), the phase of the gap function  changes very rapidly across G-X. This follows from anti-phase pairing between $d_{xy}$-electrons on first- and second-neighbor bonds.\cite{SM}  (ii) Figure \ref{fig:pip}(b) shows a gap minimum at $\th=0$ on the $\ga$ pocket, with $\Del_{\rm min}/\Del_{\rm max}< 1/100$. The near-node behavior can be ascribed to the proximity to the vHS on the zone boundary known previously,\cite{QHWepl} but the SOC reduces the amplitude (at $\th=0$) further by more than one order of magnitude, in comparison to the gap (dashed line) when SOC is artificially set to zero.\cite{note} (iii) On the $\ga$ pocket, the gap is also small at $\th=45^{\rm o}$ (or along G-M), which would be close to the gap maximum without SOC. This feature is related to the fact that the $d_{xy}$-weight is missing on the Fermi pocket along G-M (see Fig.\ref{fig:band}), whereas the dominant pairing component involves $d_{xy}$-orbital.\cite{SM} (iv) SOC also induces sizable and anisotropic gaps on the $\al$ and $\bt$ pockets, significantly larger than that without SOC.\cite{QHWepl} \\

We calculate various properties of the SC state using the FRG-derived mean field theory,\cite{SM} and compare to the experimental data. No other tuning parameters are invoked regarding the gap structure.\cite{active_ga1} The results are presented in Fig.\ref{fig:prop}. In the experimental regime, our gap structure behaves effectively nodal, and could in fact fit the data better than that in the $d$-wave case suggested in Ref.\cite{kappa3}. The details are as follows.

In Fig.\ref{fig:prop}(a), we show the specific heat in our chiral $p$-wave case (solid line), which is in excellent agreement with the experimental data (symbols) extracted from Ref.\cite{cve1}, both in the quasi-linear behavior below $T_c/2$ and the jump at $T_c$. In comparison, the $d$-wave fit (dashed line) is much poorer in both aspects.

In Fig.\ref{fig:prop}(b) we show the superfluid density $\rho$. The experimental data (symbols) are extracted from Ref.\cite{sfde} where $T_c=1.39$K. We estimate the elastic scattering rate $\zeta$ from nonmagnetic impurities in the experimental situation as,\cite{AGe1,AGe2}
\eqa \ln (T_{c0}/T_c) = \Psi(1/2+\zeta/2\pi T_c ) - \Psi(1/2), \label{eq:digma}\eea
where $\Psi(x)$ is the digamma function, $T_{c0}=1.5K$ is assumed to be the transition temperature in the disorder-free material. We get $\zeta/T_c\sim 0.1$ for $T_c=1.39$ K according to Eq.\ref{eq:digma}. Using this value of $\zeta$, the result for the chiral $p$-wave (green line) deviates from the data (symbols) in view of the curvature in the intermediate temperature window. However, if we assume $\zeta/T_c=0.5$, the result (blue line) is in much better agreement with the data, suggesting that either the sample in the experiment is dirtier than the estimate according to Eq.\ref{eq:digma}, or the clean limit $T_{c0}$ might be even higher than 1.5K. In comparison, the $d$-wave fits (dashed lines) for both scattering rates deviate from the data.

The spin-lattice relaxation rate $1/T_1$ is shown in Fig.\ref{fig:prop}(c). The theoretical result in our chiral $p$-wave case (solid line) is in good agreement with the experimental data (symbols) extracted from Ref.\cite{t1te} (where $T_c=1.48K$ corresponds to $\zeta /T_c = 0.02$ via Eq.\ref{eq:digma}). Note the approximate power-law behavior $1/T_1 \propto T^3$ in the intermediate temperature regime. The $d$-wave fit (dashed line) show similar but slightly poorer agreement. The Knight shift $K_{\mu \mu}$ depends on the probed spin direction $\mu$, see Fig.\ref{fig:prop}(d). $K_{xx,yy}$ barely changes, while $K_{zz}$ is suppressed below $T_c$. This is because our pairing function is dominated by the triplet component with its $d$-vector along $z$,\cite{SM} so that the spin of the Cooper pair lies in the plane and can response, in the linear limit, to weak in-plane (out-of-plane) field without (by) pair breaking. In experiment, $K_{zz}$ is also unchanged below $T_c$, and this is explained by the fact that the experimental field is large enough to rotate the $d$-vector, given the small energy gap. \cite{ksru1, Takamatsu}

Figure \ref{fig:prop}(e) shows the calculated $\kappa/T$ (lines) versus $T$ with $\zeta / T_c = 0.26$, along with the experimental data (symbols) with $T_c=1.2$K.\cite{kappa3} We find our chiral $p$-wave result (solid line) agrees to the data much better than the $d$-wave case (dashed line), in view of the curvature in the intermediate temperature window.
Figure \ref{fig:prop}(f) shows the calculated $\ka/T$ versus $\zeta$ (lines) at the fixed low temperature $T=T_0=T_c/30$, compared to the experimental data (symbols) from Refs.\cite{kappa1, kappa2, kappa3}. We see our chiral $p$-wave case (solid line) fits the data very well, including the universal behavior \cite{kappa-d} at $\zeta/T_0=30\zeta/T_c\gg 1$, and the mild decrease near and below $\zeta/T_c=0.4$. In contrast, in the $d$-wave case (dashed line) $\ka/T$ increases monotonically with decreasing $\zeta$, although it also shows universal behavior on the large-$\zeta$ side. (Note the eventual rise as $\zeta/T_0\ra 0$ is beyond the realm of the theory of universal conductance even for the $d$-wave case,\cite{kappa-d, SM} but in both cases can be explained by a Boltzman equation for well-defined quasiparticles, which predicts $\ka/T_0\propto 1/\zeta$. On the other hand, we have normalized the numerical $\ka/T$ by $\eps_n$, the value of $\ka/T$ with $T=T_0$, $\zeta=0.6T_c$, and zero gap. This leaves the relative size of $\ka/T$ in the $p$- and $d$-wave cases unambiguous.) Therefore, the experimental data, rather than implying $d$-wave pairing, actually supports a gap structure with various gap minima on the three Fermi pockets, as in our chiral $p$-wave case. This is supported by further discussions in Ref.\cite{SM} Of course, if the probing temperature $T_0$ is reduced further, so that $T_0\ll \Del_{\rm min}$, $\ka/T$ is eventually suppressed.\cite{SM} At this stage the $d$-wave and chiral $p$-wave behave most differently. Measurement at such low temperatures is important to close the issue, but might be a challenge in experiment. \\

{\bf Summary and remarks}: We studied the superconductivity of Sr$_2$RuO$_4$ by the state-of-art SMFRG.
We find that chiral $p$-wave pairing is dominant, but SOC induces deep near-nodes on the $\ga$ pocket and also sizable and anisotropic gaps on the $\al$ and $\bt$ pockets. The microscopic theory is in excellent agreement with experiments, resolving the outstanding puzzle between the $d$-wave-like feature in thermal measurements and the chiral $p$-wave superconductivity revealed in NMR and Kerr effect experiments.

Remarkably, the simultaneous presence of deepest near-nodes along G-X and less deep ones along G-M (both on the $\ga$ pocket in our case) is exactly the gap structure speculated to explain the systematic angle-dependent specific heat under inplane as well as conical magnetic fields in Ref.\cite{c-angle}, where the near-nodes along G-M were assumed (but do not have) to be on the $\al$ and $\bt$ pockets. The near-nodes may also be an important factor to reduce the spontaneous edge current (not detected so far\cite{exp_edge}) at finite temperatures and under impurity scattering.\cite{huang_edge, Lederer, suzuki} We leave these as future topics.

\acknowledgments{The project was supported by the National Key Research and Development Program of China (under Grant No. 2016YFA0300401), the National Basic Research Program of China by MOST (under Grant No. 2014CB921203), and the National Natural Science Foundation of China (under Grant Nos.11604168, 11574134, 11674278, and 11404383). FCZ also acknowledges the support by the Strategic Priority Research Program of the Chinese Academy of Sciences (under Grant No. XDB28000000). WSW also acknowledges the support by K. C. Wong Magna Fundation of Ningbo University.}\\


\begin{references}
\bibitem{note} For close comparison, the gaps with and without SOC are plot on the same momentum contour in the $\ga$ band.
\end{references}
\end{document}